\documentclass{Interspeech2024}

\interspeechcameraready

\title{How to Learn a New Language? An Efficient Solution for Self-Supervised Learning Models Unseen Languages Adaption in Low-Resource Scenario}

\name[affiliation={1}]{Shih-Heng Wang}{}
\name[affiliation={1}]{Zih-Ching Chen}{}
\name[affiliation={2}]{Jiatong Shi}{}
\name[affiliation={1}]{Ming-To Chuang}{}
\name[affiliation={1}]{Guan-Ting Lin}{}
\name[affiliation={1}]{Kuan-Po Huang}{}
\name[affiliation={3}]{David Harwath}{}
\name[affiliation={4}]{Shang-Wen Li}{}
\name[affiliation={1}]{Hung-yi Lee}{}

\address{
  $^1$National Taiwan University, Taiwan
  $^2$Carnegie Mellon University, US \\
  $^3$The University of Texas at Austin, US
  $^4$FAIR, US}
  
\email{}


\usepackage{mathtools}
\usepackage{amsmath}
\usepackage{float}
\newcommand{\argmin}{\mathop{\mathrm{argmin}}} 

\usepackage{multirow}
\usepackage{multicol}
\usepackage{color, colortbl}
\usepackage{xcolor}
\usepackage{comment}
\usepackage{cite}
\usepackage{array}
\usepackage{subcaption}
\usepackage{threeparttable}
\usepackage{algorithm}
\usepackage{algpseudocode}
\usepackage{hyperref}
\usepackage{tikz}
\usepackage{xcolor}
\usepackage[normalem]{ulem} % for strikethrough

\colorlet{full}{green!10}
\colorlet{cmn}{teal!10}
\colorlet{FOMAML}{purple!10}
\colorlet{MTL}{purple!10}
\colorlet{MTL*}{purple!10}
\colorlet{multi}{purple!10}
\colorlet{euro}{orange!10}
\begin{document}

\maketitle

\begin{abstract}
The utilization of speech Self-Supervised Learning (SSL) models achieves impressive performance on Automatic Speech Recognition (ASR). However, in low-resource language ASR, they encounter the domain mismatch problem between pre-trained and low-resource languages. Typical solutions like fine-tuning the SSL model suffer from high computation costs while using frozen SSL models as feature extractors comes with poor performance.
To handle these issues, we extend a conventional efficient fine-tuning scheme based on the adapter. We add an extra intermediate adaptation to warm up the adapter and downstream model initialization. Remarkably, we update only 1-5\% of the total model parameters to achieve the adaptation. Experimental results on the ML-SUPERB dataset show that our solution outperforms conventional efficient fine-tuning. It achieves up to a 28\% relative improvement in the Character/Phoneme error rate when adapting to unseen languages.
\end{abstract}

\section{Introduction}
\label{sec: intro}

% Paragraph 1: Introduce SSL models (Monolingual and Multilingual) and mention low resource ASR
Self-Supervised Learning models (SSL models)~\cite{Wu2021CrossLingualTF, adams2019massively, Hou2020LargeScaleEM, Khare2021LowRA, yadav-sitaram-2022-survey, 9023195} pre-trained with speech-only data have achieved significant improvements for Automatic Speech Recognition (ASR) in mainstream languages. \cite{baevski2020wav2vec, hsu2021hubert, babu22_interspeech, chung2021w2v, Chen_2022, Mohamed_2022, shi2024multiresolution}.
However, employing SSL models on low-resource language ASR may encounter the problem of domain mismatch between pre-trained and low-resource languages \cite{seth2022analyzing}.
Since SSL models are mostly pre-trained with high-resource languages, like English, they may not generalize well on those low-resource languages~\cite{shi2023mlsuperb, Liu2023StudyingTI, singh2023novel}. 
% Paragraph 2: points out current problem of low resource ASR 

We aim to solve the mismatch problem when employing SSL models on low-resource language ASR. Typically, there are two applicable approaches: fine-tuning the SSL models with the low-resource language data and employing the SSL models as feature extractors for subsequent downstream models.~\cite{shi2023mlsuperb, yang21c_interspeech}. 
Fine-tuning the SSL model with the downstream model leads to better performance but suffers from high computation costs. Furthermore, the amount of target low-resource language data is extremely insufficient for large-scale model training. Therefore, it may lead to sub-optimal transfer performance due to over-parameterization. On the other hand, utilizing frozen SSL models as feature extractors could be a lower-cost option. However, it usually comes with poor performance, especially when the target low-resource language is unseen to the SSL model. Apart from these solutions, Adapters~\cite{udupa24_interspeech, lester-etal-2021-power, houlsby2019parameter,fu-etal-2022-adapterbias, li-liang-2021-prefix,  zaken2021bitfit, chen2023adapter, fu-etal-2022-adapterbias,  chen2023chapter,thomas2022efficient}, which are lightweight modules inserted in the pre-trained model, could be a preferable solution since it only fine-tunes limited amount of inserted parameters, achieving a balance between performance and computation cost. However, they could still encounter difficulties in transferring to an unseen language in low-resource scenarios due to domain mismatch.

% (1) Large Model -> Current paradigm freeze SSL and only fine-tune downstream model -> low cost but bad performance, when the language is unseen to SSL models, the performance is very bad

% (2) Fine-tune SSL models may lead to better performance but with high computation and storage cost.
% when the amount of pair data is limited (low-resource)  -> Bad transfer performance due to over-parameterized

% Paragraph 3: 

% Introduce Existing works, and their limitation, Need to mention adapter (Focus on adaptation paper and adapter, end with 'https://arxiv.org/pdf/2302.01496.pdf')
\begin{figure}[]
\centering
    \includegraphics[width=1\linewidth, height=4.5cm]{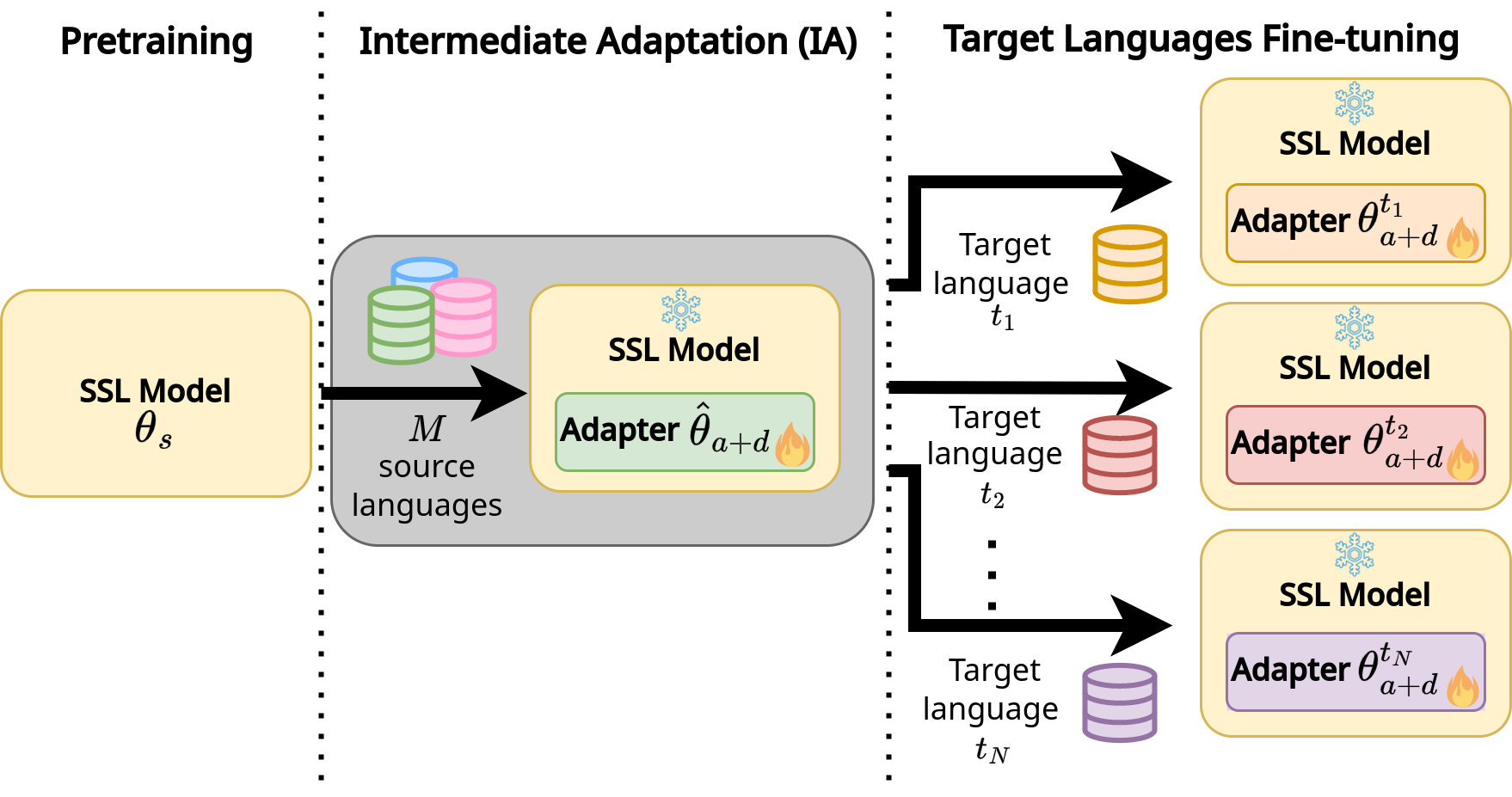}
    \caption{Pipeline of our solution. Before fine-tuning the adapter and downstream model (omitted in the figure) to each target language, we warm up them with Intermediate Adaptation.  }
    \label{fig:pipeline}
\end{figure}
Some existing works aim to solve the domain mismatch problem. For example, \cite{kessler2022adapter} introduces continual pre-training with the adapter. Nevertheless, its need for large-scale labeled data makes it unsuitable for low-resource scenarios. \cite{hou2021exploiting} leverages multiple adapters trained on high-source languages to enhance performance on low-resource languages. However, it requires multiple adapters at the same time during training and inference, becoming impractical as the number of languages increases. \cite{10096330} presents a recipe to efficiently adapt models to a different domain. Nevertheless, they did not explore the cross-language adaptation.
% Paragraph 4: Introduce our solution

%     4-1 We want good results and low cost at the same time -> 
    
%     4-2 Introduce pipeline (mentioned Fig1), Follow "https://arxiv.org/pdf/2302.01496.pdf" paradigm to use adapter and extra intermediate stage before fine-tuning
    
%     4-3 In our scenarios, problem will be (1) which language to use? (2) What algorithm to employ in IA
    
%     4-4 Briefly introduce how we select intermediate languages and comparison between MTL and MAML
To deal with the domain mismatch problem, we provide an efficient solution.
% Furthermore, we would like to keep our solution efficient. 
Figure~\ref{fig:pipeline} shows the general pipeline of our solution. Our solution utilizes adapters to keep the computation cost low. 
To facilitate the adaptation, we add an \textbf{Intermediate Adaptation} (IA) before fine-tuning the adapter and downstream model on low-resource target languages. IA serves as a bridge between pre-trained languages and unseen target languages. During the IA, we utilize various adaptation algorithms to warm up the adapter and downstream model with high-resource source languages. These high-resource languages are selected to optimize the model's transferability to unseen target languages. After IA, we can derive an enhanced initialization and perform Parmeter-Efficient Fine-tuning (PEFT)~\cite{peft} to fine-tune on each target language.
% In other words, IA serves as a bridge between pre-trained languages and unseen target languages. 
% Third, we further explore the adaptation algorithm employed in the IA. 
% In conclusion, based on \cite{10096330}, we provide a full recipe to efficiently adapt SSL models to unseen target languages.
Experimental results on the ML-SUPERB dataset~\cite{shi2023mlsuperb} demonstrate that our solution outperforms conventional efficient fine-tuning. In the best case, it achieves up to a 28\% relative improvement in the Character/Phoneme error rate when adapting to unseen target languages. Furthermore, we provide analysis for different SSL models and our proposed source language selection methods.

% Paragraph 5: Briefly introduce the result and give temporary conclusion

\section{Methodology}
\label{sec:method}
We focus on effectively adapting SSL models to each unseen target language in low-resource scenarios. As illustrated in Figure~\ref{fig:pipeline}, before fine-tuning on each target language, we add an extra \textbf{Intermediate Adaptation} (IA) step. IA warms up the adapter and downstream model with source languages to facilitate adaptation to each unseen target language. After IA, we obtain an enhanced adapter and downstream model initialization (the green module in Figure~\ref{fig:pipeline}). With this initialization, we apply PEFT to fine-tune the adapter and the downstream model on each target language (the red, orange, and purple adapter in Figure~\ref{fig:pipeline}). It is important to note that the SSL model is frozen all the time, making our solution low-cost.

% \subsection{
Here, we would like to give a general formulation of IA. Given model initialization $\theta = \theta_s \cup \theta_a \cup \theta_d$ (including frozen SSL model $\theta_s$, randomly initialized adapter $\theta_a$ and downstream model $\theta_d$) and source languages $\mathcal{S} = \{ s_1, s_2, ..., s_M \}$ ($\mathcal{S}$ is a set with $M$ languages), the objective of the IA is to find an enhanced initialization $\hat{\theta}_a \cup \hat{\theta}_d $ (abbreviated into $\hat{\theta}_{a+d}$ for better readability, \textbf{$\hat{}$} means the parameter is warmed up). In other words, IA can be formulated as

     \begin{equation} \scalebox{0.95}[1]\small
       \hat{\theta}_{a+d} = \mathrm{IA}(\theta, \mathcal{S}),
       \label{eq:IA}
    \end{equation}
    where $\hat{\theta}_{a+d}$ is the initialization for performing PEFT on each target language in $\mathcal{T} = \{ t_1, t_2, ..., t_N \}$ ($\mathcal{T}$ is a set with $N$ languages). The concrete form of IA($\cdot$) will be provided in the Section~\ref{subsec:algorithm}. 
    % Since only the $\theta_a \cup \theta_d$ (abbreviated into $\theta_{a+d}$) are updated during IA, our solution maintains low computation and storage costs. 
    To obtain $\hat{\theta}_{a+d}$ with the best adaptation result on $\mathcal{T}$, there are two problems to address: (1) What kinds of \textbf{source languages $\mathcal{S}$} (see Section~\ref{subsec:source_lang_selection}) and (2) \textbf{adaptation algorithms}(see Section~\ref{subsec:algorithm}) can best facilitate the adaptation to unseen target languages $\mathcal{T}$?
\subsection{Source Language Selection}
\label{subsec:source_lang_selection}
\begin{figure}[t]
\centering
    \includegraphics[width=0.9\linewidth, height=4.5cm]{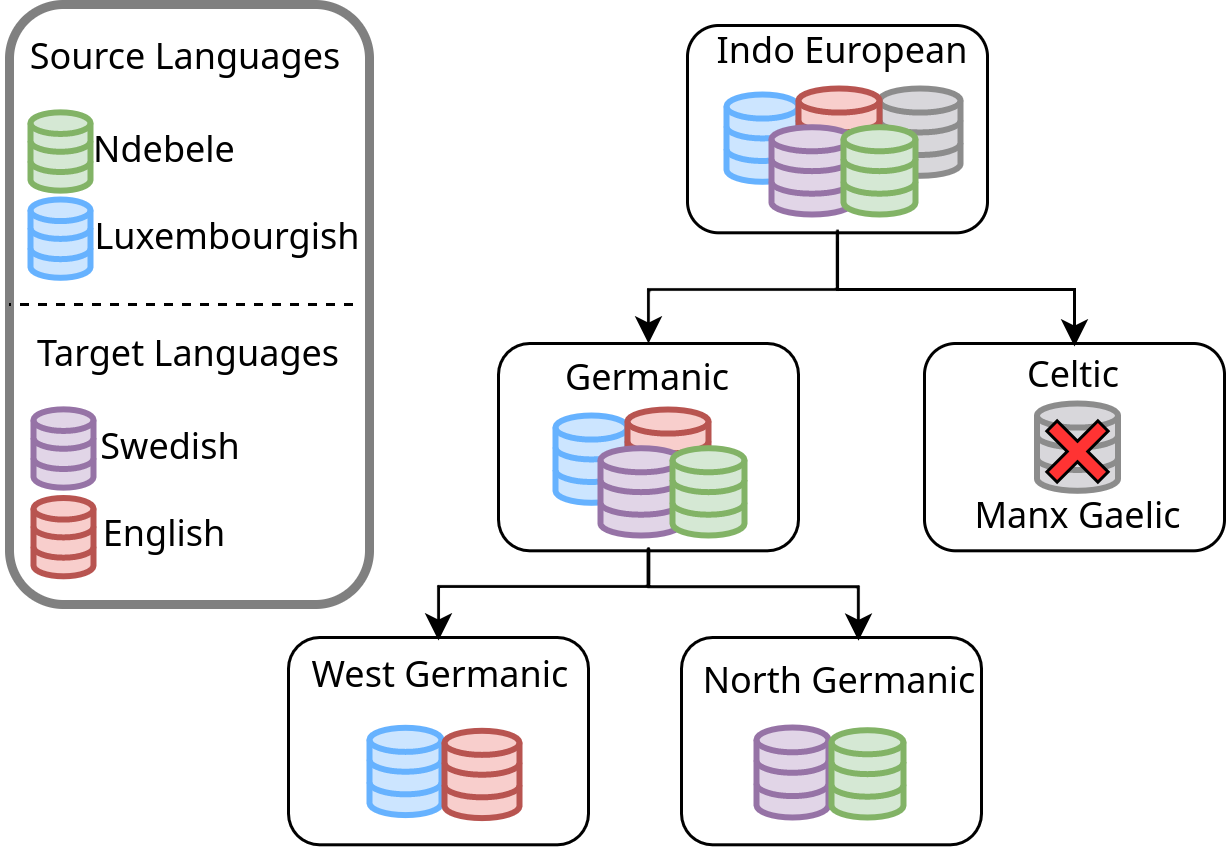}
    \caption{Our source language selection process with the linguistic tree. Based on the topology of the example linguistic tree, we pick ``Luxembourgish" and ``Ndebele" instead of ``Manx Gaelic" as source languages for IA because they are linguistically closer to our target languages ``English" and ``Swedish". }
    \label{fig:tree}
\end{figure}

To identify source languages beneficial for adaptation to unseen target languages $\mathcal{T}$, we use the linguistic knowledge based on a linguistic tree~\cite{li-etal-2022-zero}. As illustrated in Figure~\ref{fig:tree}, we select source languages (``Luxembourgish", 
 ``Ndebele") linguistically close to target languages (``English", ``Swedish")\footnote{We use well-known languages as examples for better readability. In low-resource language cases, we can still employ the same method.}, as they might share some acoustic traits~\cite{Wu2021CrossLingualTF}. In other words, we assume that warming up $\theta_{a+d}$ on languages similar to $\mathcal{T}$ to get $\hat{\theta}_{a+d}$ may facilitate the final adaptation result on $\mathcal{T}$.
% Comparing to data-driven approaches, our method, which injects knowledge using linguistic similar languages, is more explainable.

The detailed implementation of selecting source languages $\mathcal{S}$ is explained here. Given target languages $\mathcal{T}$, we traverse the linguistic tree, exclude languages in $\mathcal{T}$, and select the top $M$ most linguistically similar languages as $\mathcal{S}$. We define the linguistic-similarity function Sim($\cdot$) to $\mathcal{T}$ using Lowest Common Ancestor (LCA):

    \begin{equation}
    \scalebox{0.9}[1]{$\displaystyle
        \textrm{Sim}(l, \mathcal{T}) = \sum\limits_{j=1}^N D(\mathrm{LCA}(l, t_j)),
    $}
    \label{eq:tree}
    \end{equation}

    where $l$ is the language in the linguistic tree, and $D$ computes the depth of a node in the tree. A higher Sim($l$, $\mathcal{T}$) value implies that the language $l$ is linguistically closer to $\mathcal{T}$ in the tree. For example, in Figure~\ref{fig:tree}, we pick the blue (``Luxembourgish") and green (``Ndebele") ones instead of the gray one (``Manx Gaelic") because their depths of LCA (Germanic) to target languages are deeper than that of the gray one (Indo European).

\subsection{Adaptation Algorithm}
\label{subsec:algorithm}
    Appropriate selection of an adaptation algorithm in IA also has a huge impact on the final adaptation result. The adaptation algorithm finds $\hat{\theta}_{a+d}$ based on the $\theta $ and $\mathcal{S}$ (see Eq. ~\eqref{eq:IA}).  
    To find the adaptation algorithm with the best adaptation result, we explore two prominent algorithms:
    
    \noindent \textbf{Multitask Learning (MTL)}:
    Multitask Learning (MTL) seeks to optimize the initialization $\theta$ across all source languages $\mathcal{S}$ simultaneously to get $\hat{\theta}_{a+d}$. In other words, following the IA general form (see Eq.~\eqref{eq:IA}), the optimization objective of MTL can be formulated as: 
    \begin{equation} \scalebox{0.95}[1]\small
        \hat{\theta}_{a+d} = \mathrm{IA}(\theta, \mathcal{S}) = \argmin_{\theta_a \cup \theta_d}\sum_{i=1}^{M}\limits L(\theta, s_i),
    \end{equation}
    where $L$ denotes the ASR loss on each source language.

    \noindent \textbf{Model-Agnostic Meta-Learning (MAML)}:
    MAML\cite{finn2017modelagnostic} is a commonly adopted algorithm in few-shot adaptation scenarios. Unlike MTL, MAML adopts a bi-level optimization process, which includes the inner and outer loop. Following the IA general form (see Eq.~\eqref{eq:IA}), the optimization objective of MAML can be formulated as: 
    \begin{equation} \scalebox{0.95}[1]\small
        \hat{\theta}_{a+d} = \mathrm{IA}(\theta, \mathcal{S}) = \mathrm{MAML}(\theta, \mathcal{S}),
    \end{equation}
    where the MAML function is defined at Alg.~\ref{alg:MAML}. As shown in Alg.~\ref{alg:MAML}, in the while loop, we first sample a batch of data $B_i$ from a source language $s_i$ $\sim$ $\mathcal{S}$. Next, we split the $B_i$ into support set $B_{i}^{s}$ and query set $B_{i}^{q}$. In the inner loop, we derive the language-specific model $\theta_i'$ with $B_{i}^{s}$. Last, we calculate the gradient using $\theta_i'$ and $B_{i}^{q}$ to update $\theta$. Until the ASR loss $L$ of the outer loop converges, we adopt the $\theta_{a+d}$ as $\hat{\theta}_{a+d}$.
    \begin{algorithm}
    \caption{ MAML($\theta$, $\mathcal{S}$)}
    \begin{algorithmic}[1]
    \Require $\mathcal{S}$: Source languages
    \Require $\theta$: model parameters
    \Require $\alpha, \beta$: inner \& outer loop learning rate
    \Require $f_{\theta}$: model function parametrized by ${\theta}$
     \While{not done}
        \State Sample a batch of data $B_i$ from language $s_i \sim \mathcal{S}$
        \State Split the batch into support set $B_{i}^{s}$ and query set $B_{i}^{q}$ 
        \ForAll{$B_{i}^{s}$}
            % \State Evaluate $\nabla_\theta \mathcal{L}_{B_{i}^{s}} (f_\theta)$ 
            \State Compute adapted parameters with gradient descent: 
              \State $\theta_i' = \theta - \alpha \nabla_\theta L_{B_{i}^{s}} (f_\theta)$
        \EndFor
        \State Update $\theta \gets \theta - \beta \nabla_\theta L_{B_{i}^{q}} (f_{\theta_i'})$
    \EndWhile
    \end{algorithmic}
    \label{alg:MAML}
    \end{algorithm}
\subsection{Target Languages Fine-tuning}
    After deriving $\hat{\theta}_{a+d}$ with IA (see Figure~\ref{fig:pipeline}), we fine-tune $\hat{\theta}_{a+d}$ to each target language. Specifically, Figure~\ref{fig:pipeline} illustrates that we fine-tune $\hat{\theta}_{a+d}$ to each target language $\{ t_1, t_2, ..., t_N \}$ to get $\{\theta_{a+d}^{t_1}, \theta_{a+d}^{t_2}, ..., \theta_{a+d}^{t_N} \}$ (a set of $N$ adapter and downstream model parameters). 

\section{Experimental Setups}

\label{sec:experiment}
% In this section, we introduce our experimental setup. 

\subsection{Dataset}

    We evaluate our solution using ML-SUPERB \cite{shi2023mlsuperb}, a benchmark for multilingual ASR with speech SSL models. ML-SUPERB is supported by 143 languages. For each language, ML-SUPERB provides 10-minute and 1-hour settings. The duration means the training data size employed in fine-tuning for each language. For evaluation metrics, we follow the ML-SUPERB settings to report the Character/Phoneme Error Rate (CER/PER).

    \subsection{Source and Target Languages }
    \label{subsec:source_target}
    We use the ML-SUPERB dataset to construct the source and target language sets $\mathcal{S}$ and $\mathcal{T}$. For target languages $\mathcal{T}$, we build two target language sets: the \textbf{Seen Set} and the \textbf{Unseen Set}. Each of them has its corresponding source languages $\mathcal{S}$ using our proposed method (see Section~\ref{subsec:source_lang_selection}). For the amount of training data, we use 10-minute and 1-hour settings for target languages while using the 1-hour setting for 
 source languages.
    
    Table~\ref{tab: source_target_set} lists the source and target languages of the two sets. The explanation of the two sets are shown below:
    
    \noindent\textbf{Seen Set}: This set is derived from MLSUPERB's Monolingual Track, including 9 widely used languages (see Table~\ref{tab: source_target_set} (I)). These languages are seen by some SSL models we use in the experiment. This set is intended for direct comparison with MLSUPERB results and rapid concept validation. 
    
    \noindent\textbf{Unseen Set}. This set is derived from the ML-SUPERB Multilingual Track, including 20 endangered languages (see Table~\ref{tab: source_target_set} (II)). This set evaluates the model's adaptability to unseen languages, given that our SSL models did not previously see these languages during pre-training.
    
    \begin{table}[]
    \centering
    \caption{Detailed source \& target languages of the Seen \& Unseen Set (using ISO-639 code)}
    % \caption{Percentage of updated parameters}
    % \scalebox{0.9}{
% \begin{threeparttable}
% \begin{tabular}{l|>{\centering\arraybackslash}m{3cm}|>{\centering\arraybackslash}m{3cm}}
% \toprule
%   & \textbf{Seen Set} & \textbf{Unseen Set} \\
% \midrule 
% \shortstack{Source \\Languages } & \shortstack{ltz, nor, spa, por, oci, \\ nld, glg, cat, ast, afr} & \shortstack{xho, ven, ssw, sot, sna, \\ nso, nbl, nya, lug, kin} \\
% \midrule
% \shortstack{Target\\ Languages} & \shortstack{eng, fra, deu, rus, swa, \\  swe, jpn, cmn, xty} & \shortstack{dan, epo, frr, tur, lit, \\ srp, vie, tok, kaz, umb, \\ zul, bos, ful, ceb, luo, \\ kea, sun, tsn, tso, tos} \\
% \bottomrule
% \end{tabular}
% \end{threeparttable}
% }
\scalebox{0.9}{
\begin{threeparttable}
\begin{tabular}{>{\centering\arraybackslash}m{2cm}|>{\centering\arraybackslash}m{2.8cm}|>{\centering\arraybackslash}m{2.8cm}}
\toprule
\textbf{Target Language Set} & \textbf{Source Languages $M$=10}  & \textbf{Target Languages}  \\ 
\midrule 
\midrule
(I) Seen Set & \shortstack{ltz, nor, spa, por, oci, \\ nld, glg, cat, ast, afr} &  \shortstack{eng, fra, deu, rus, swa, \\  swe, jpn, cmn, xty}  \\ 
\midrule
(II) Unseen Set  & \shortstack{xho, ven, ssw, sot, sna, \\ nso, nbl, nya, lug, kin} &   \shortstack{dan, epo, frr, tur, lit, \\ srp, vie, tok, kaz, umb, \\ zul, bos, ful, ceb, luo, \\ kea, sun, tsn, tso, tos}  \\
\bottomrule
\end{tabular}
\end{threeparttable}
}
    \label{tab: source_target_set}
    \end{table}
     % \noindent\textbf{Mainstream Languages Set}: {ltz,nor,spa,por,oci,nld,glg,cat,ast,afr}
    
    % \noindent\textbf{Reserve Languages Set}: {xho,ven,ssw,sot,sna,nso,nbl,nya,lug,kin}
    % In essence, the cumulative volume of data $\mathcal{S}$ used in the priming stage is less than 40 hours.

\subsection{Model Configuration \& Hyperparameter}
\label{subsec:model}

    % \subsubsection{SSL Model, Adapters, and Downstream Model}
    % Here we discuss our model configuration and hyperparameter settings.
    % \newline
    \noindent\textbf{SSL Model ($\theta_s$)}. We employ three onboard pre-trained SSL models: HuBERT-base \cite{hsu2021hubert}, mHuBERT-base\cite{lee2022textless}, and XLSR-128\cite{babu22_interspeech}. 
    These models encompass various traits, including monolingual / multilingual, base / large size.
    
    \noindent\textbf{Adapters ($\theta_a$)}. Adapters are lightweight modules inserted in neural networks that enable task-specific adaptations without modifying the original model's parameters. Adapter modules are added to the second feed-forward layers of transformer layers in the SSL model, operating independently from the downstream model. The adapter implementation strictly adheres to the methodology in \cite{chen2023adapter}, specifically adopting the Houlsby adapter~\cite{houlsby2019parameter}. The bottleneck is set to 32.
    
    \noindent\textbf{Downstream Model ($\theta_d$)}. The downstream model ($\theta_d$) adopts a transformer architecture with the connectionist temporal classification (CTC) objective as outlined in the ML-SUPERB~\cite{shi2023mlsuperb}. Both the model's architecture and hyper-parameters follow the specifications in ML-SUPERB for a fair comparison. It is important to note that we reinitialize the CTC head of the downstream model after IA because the characters set of source languages are different from that of target languages. 
    % That is, $\hat{\theta}_{a+d}$ and $\{\theta_{a+d}^{t_1}, \theta_{a+d}^{t_2}, ..., \theta_{a+d}^{t_N} \}$ all have different CTC heads.
    
    Table~\ref{tab: Param} shows the percentages of the $\theta_a$ and $\theta_d$ parameters.
    Note that we always freeze the $\theta_s$ to make our solution low cost. 
    
    % \noindent \textbf{Implementation Detail:} Most implementation details of our work follow the default settings in \cite{shi2023mlsuperb} using ESPnet. The architecture of the downstream model, training details, and evaluation metrics are consistent with \cite{shi2023mlsuperb}. 

    \noindent\textbf{Adaptation Algorithms} (see Section~\ref{subsec:algorithm}). For the MAML, we adopt the first-order version (FOMAML) to save computation costs. For the MAML hyper-parameters, we set the $\alpha$ = 0.001, $\beta$ = 0.0001, inner step = 1, and update the model using SGD in the inner loop and Adam in the outer loop.  For the MTL, we set the learning rate to 1e-4, using Adam optimizer.

    % To offer a comprehensive analysis, our study encompasses scenarios \textbf{without the use of Adapters} in Sec~\ref{sec:prime_parameter}. In such cases, $\theta_a$ remains uninitialized, and solely $\theta_d$ instead of $\theta_a \cup \theta_d$ undergo priming and fine-tuning.

    \begin{table}[]
    \centering
        % \captionsetup{justification=centering}
        % \captionsetup{font=footnotesize}
    \caption{Percentage of tunable parameters. 100\% = $\theta_s \cup \theta_a \cup \theta_d$}
    % \caption{Percentage of updated parameters}
    \scalebox{0.9}{\begin{threeparttable}

% \caption{Percentage of updated parameters}
\begin{tabular}{c|cc}

\toprule
\multirow{2}{*}{SSL} & \multicolumn{2}{c}{Tunable Parameters } \\ \cmidrule{2-3} 
                     & \multicolumn{1}{c|}{Downstream Model $\theta_d$} & Adapter $\theta_a$ \\ \midrule
HuBERT/mHuBERT(95M)     & \multicolumn{1}{c|}{4.6M (4.6\%)}       & 0.7M (0.7\%)    \\ \midrule
XLSR-128(317M)             & \multicolumn{1}{c|}{4.6M (1.4\%)}       & 1.6M (0.5\%)    \\ \bottomrule
\end{tabular}
% \caption*{\fontsize{9pt}{11pt}\selectfont``w/o A'' and ``w/ A'' represents cases without and with adapter.}
\end{threeparttable}
}
    \label{tab: Param}
    % \vspace{-20pt}
    \end{table}

\subsection{Baselines}
\label{subsec:baselines}
     To prove the effectiveness of our pipeline, we adopt four common fine-tuning methods as our baselines: \textit{Full FT}, \textit{Freeze FT}, \textit{PEFT}, and \textit{Source \& Target ($\mathcal{S}$\&$\mathcal{T}$)-MTL}. 
    % To ensure a fair comparison between each baseline, the tuned parameters (whether in the priming stage or fine-tuning stage) are consistently set to $\theta_a \cup \theta_d$.
    
    \noindent\textbf{\textit{Full FT ($\theta_{s+d}$):}} This is the SSL model fine-tuning baseline, where we fine-tune the $\theta_{s+d}$ on each target language in $\mathcal{T}$, without initializing $\theta_a$ and IA. This method serves as the strong baseline due to its high computational and storage costs.
    
    \noindent\textbf{\textit{Freeze FT ($\theta_{d}$):}} This is the widely-used fine-tuning baseline, where we freeze $\theta_s$ and fine-tune $\theta_d$ on each target language in $\mathcal{T}$, without initializing $\theta_a$ and IA. This method is the default setting of ML-SUPERB.

    \noindent\textbf{\textit{PEFT ($\theta_{a+d}$):}} This is the PEFT baseline. It freezes $\theta_s$ and fine-tunes $\theta_{a+d}$ on each target language in$\mathcal{T}$, without IA.

    \noindent\textbf{\textit{$\mathcal{S}$\&$\mathcal{T}$-MTL ($\theta_{a+d}$):}} This is the baseline using the same amount of training data without two-stage training like IA. It involves jointly fine-tuning $\theta_{a+d}$ on both the $\mathcal{S}$ and $\mathcal{T}$ to get a multilingual model without further fine-tuning on each target language in $\mathcal{T}$.

\section{Result \& Analysis}
\subsection{Main Result}

Table~\ref{tab: MR} presents the results of our IA variants and baselines from the Seen Set and the Unseen Set. Remarkably, two IA variants (\textit{IA-MAML}, \textit{IA-MTL}) consistently outperform other baselines (\textit{Freeze-FT}, \textit{PEFT}, \textit{$\mathcal{S}$\&$\mathcal{T}$-MTL}) on both sets, while \textit{IA-MTL} slightly outperforms \textit{IA-MAML}. In the 10-minute and 1-hour setting, IA variants achieve substantial improvements. Compared with \textit{PEFT} baseline, IA variants achieve up to 28\% and 20\% relative improvement. Impressively, when compared to a strong baseline like \textit{Full-FT}, two IA variants either surpass or match \textit{Full-FT} performance, but require less than 6\% of the tunable parameters in \textit{Full-FT}. Our results strongly support the effectiveness of IA in facilitating language adaptation.
\begin{table}[t]
    % \captionsetup{justification=centering}
    % \captionsetup{font=footnotesize}
    \caption{Character/Phoneme Error Rate (CER/PER) of Seen Set and Unseen Set. The 10min and 1h are the amount of fine-tuning data for each target language. Pink rows are IA variants while green rows are strong baselines (see Section~\ref{subsec:baselines})}
    % \vspace{-0.2cm}
    \centering
    \include{tables/Mono&Reserve_cer}
    \label{tab: MR}
    \vspace{-15pt}
    \end{table}
    
    \begin{table}[h]
        % \captionsetup{font=footnotesize}
        \caption{(a) Source language selection method comparison (b) Optimal number of source languages $M$ using our method }
        % \vspace{-0.3cm}
        \begin{subtable}{.49\linewidth}
            % \captionsetup{justification=centering}\captionsetup{font=footnotesize}
                \centering
                \caption{}
            \scalebox{0.75}{
\begin{threeparttable}
      \captionsetup{justification=raggedright}
       \captionsetup{font=footnotesize}
% \caption*{Please refer to \ref{enu: source_language_selection} and APPENDIX   for  the notation of baselines and the details of the languages used.}
% \caption*{}

\begin{tabular}{c|cc}

\toprule
\multirow{2}{*}{\begin{tabular}[c]{@{}c@{}}Source Language \\ Selection Methods\end{tabular}} & \multicolumn{2}{c}{CER / PER $\downarrow$}              \\ \cmidrule{2-3} 
                                                                                             & \multicolumn{1}{c|}{10min} & 1h      \\ \midrule
 \midrule
\multicolumn{1}{l|}{Random $M$ = 5 }                                                                                          & \multicolumn{1}{c|}{38.5}  & 31.0  \\ \midrule
\multicolumn{1}{l|}{Proposed $M$ = 5}                                                                                          & \multicolumn{1}{c|}{\textbf{35.9}}    & \textbf{29.5}  \\ \midrule \midrule
\multicolumn{1}{l|}{Random $M$ = 10 }                                                                                          & \multicolumn{1}{c|}{37.7}  & 30.6  \\ \midrule
\multicolumn{1}{l|}{Proposed $M$ = 10}                                                                                  & \multicolumn{1}{c|}{\textbf{34.8}}  & \textbf{28.9} \\ \bottomrule 
\end{tabular}
% \begin{tablenotes}[para,flushleft]    
%         \captionsize 
%         % \item{\textbf{Note.}} The notation for baselines can be found in \ref{enu: source_language_selection}
%         % \item{\textbf{Note.}} For languages used in 1.1-1.3, refer to APPENDIX.
        
%          \item Please refer to \ref{enu: source_language_selection} and APPENDIX   for  the notation of baselines and the details of the languages used.

%       \end{tablenotes}   
      % \caption*{Please refer to \ref{enu: source_language_selection} and APPENDIX   for  the notation of baselines and the details of the languages used.} 
% \caption {The notation for baselines can be found in \ref{enu: source_language_selection}}
% \caption {For languages used in 1.1-1.3 please refer to \ref{appendix} APPENDIX.}
\end{threeparttable}}
            \label{tab: source_language_selection}
        \end{subtable}
         \begin{subtable}{.49\linewidth}
            % \captionsetup{justification=centering}
            % \captionsetup{font=footnotesize}
            \centering
            \caption{}
            \scalebox{0.75}{
\begin{threeparttable}
      % \captionsetup{justification=raggedright}
      %  \captionsetup{font=footnotesize}
% \caption*{Please refer to \ref{enu: source_language_selection} and APPENDIX   for  the notation of baselines and the details of the languages used.}
% \caption*{}

\begin{tabular}{c|cc}

\toprule
\multirow{2}{*}{\begin{tabular}[c]{@{}c@{}}Source Language \\Number $M$\end{tabular}} & \multicolumn{2}{c}{CER / PER $\downarrow$}              \\ \cmidrule{2-3} 
& \multicolumn{1}{c|}{10min} & 1h      \\ \midrule
% \multicolumn{1}{l|}{$M$ = 0 }                                                                                          & \multicolumn{1}{c|}{39.6}  & 31.3  \\ \midrule
\multicolumn{1}{l|}{Proposed $M$ = 5 }                                                                                          & \multicolumn{1}{c|}{35.9}  & 29.5  \\ \midrule
\multicolumn{1}{l|}{Proposed $M$ = 10}                                                                                          & \multicolumn{1}{c|}{34.8}    & 28.9  \\ \midrule
\multicolumn{1}{l|}{Proposed $M$ = 20 }                                                                                          & \multicolumn{1}{c|}{\textbf{33.3}}  & \textbf{27.9}  \\ \midrule

\multicolumn{1}{l|}{Proposed $M$ = 50}                                                                                  & \multicolumn{1}{c|}{33.5}  & 	28.2 \\ 
% \midrule 
% \multicolumn{1}{l|}{Our $M$ = 100}                                                                                  & \multicolumn{1}{c|}{35.1}  & 28.9 \\
\bottomrule
\end{tabular}

\end{threeparttable}}
            \label{tab: M}
        \end{subtable}
    \vspace{-10pt}    
    \end{table}
\subsection{Impact of SSL Model Pre-trained Languages} 
    In this section, we discuss the adaptation result of SSL models to seen and unseen target languages (see Section~\ref{subsec:source_target}). 
    % For the readability, we divide the discussion into two groups:
    
    \noindent \textbf{HuBERT-Base \& mHuBERT:} Here, we discuss whether IA benefits the adaptation of base-size and less multilingual SSL models. HuBERT-Base~\cite{hsu2021hubert} is pre-trained on 1k hours of English data while mHuBERT-Base~\cite{lee2022textless} is pre-trained on 14k hours from three European languages. In Seen Set \& Unseen Set, most languages are unseen during the pre-training of these two models. Table~\ref{tab: MR} shows that IA does facilitate the adaptation of base size and less multilingual SSL models to unseen target languages. Our IA variants achieve at most 14.9\% relative improvement compared with \textit{PEFT} baseline. 
    
    \noindent \textbf{XLSR-128:} Here, we discuss whether IA benefits the adaptation of a large multilingual SSL model. XLSR-128~\cite{babu22_interspeech} is pre-trained on 400k hours of data in 128 languages. For XLSR-128, most languages in the Seen Set are seen, while those in the Unseen Set are mostly unseen during the pre-training. From Table~\ref{tab: MR}, we can tell that IA improves the adaptation result in both sets. To be more specific, compared with \textit{PEFT} baseline, our IA variants show at most 24.9\% and 28.2\% relative improvement on each set. This validates that IA does facilitate the adaptation of large multilingual SSL models to unseen target languages.
    
\subsection{Effectiveness of Source Languages Selection Methods}
    Source language selection is critical in IA. To validate our selection methods (see Section~\ref{subsec:source_lang_selection}), we compare our linguistic-knowledge-based method with randomly sampling source languages from the ML-SUPERB dataset. The experiments are conducted with $M$ = 5 and $M$ = 10 under the \textit{IA-MTL} default settings on Seen Set~\footnote{Source languages $M$=5 \{cat, mar, guj, kan, tam\},  $M$=10 \{nbl, ssw, ven, mal, ben, mri, sot, nep, sin, jav\}, ISO-639 code}. Table~\ref{tab: source_language_selection} shows that our proposed method outperforms random selection baselines in both settings, validating the effectiveness of our method. 
    
    Also, we provide experiments for the optimal number $M$ of source languages. Table~\ref{tab: M} illustrates the trend $M$ = \{5, 10, 20, 50\}. The experiment result indicates that $M = 20$ achieves the best performance. Furthermore, we see no improvement when we increase $M$ to 50. This suggests that languages chosen later, which are less linguistically similar to our target languages, contribute less to the adaptation to unseen target languages. Note that using more source languages takes more epochs for the model to converge in IA. Therefore, we set $M$ = 10 instead of 20 as our default setting due to the computational constraint.

\section{Conclusion \& Limitation}
\label{sec:conclusion}

In this work, we propose an efficient solution for adapting SSL models to unseen language in low-resource scenarios. Our solution adds an extra Intermediate Adaptation (IA) to warm up the adapter and downstream model initialization. With this enhanced initialization, the model can adapt to unseen target languages more easily. In our low-cost solution, only 1-5\% of the total model parameters are modified to adapt to each language. Experiment results on the ML-SUPERB dataset show that our solution achieves up to a 28\% relative improvement in CER/PER over conventional efficient fine-tuning. Additionally, our results validate the effectiveness of our source language selection method and our configuration of tunable parameters. Overall, our efficient solution contributes to the employment of SSL models on low-resource language ASR. For the limitation of our work, our source language selection method needs to know the target language set beforehand. Also, we lack explorations of second-order MAML and other types of adapters.

% \section{Acknowledgements}
% Acknowledgement should only be included in the camera-ready version, not in the version submitted for review.
% The 5th page is reserved exclusively for \red{acknowledgements} and  references. No other content must appear on the 5th page. Appendices, if any, must be within the first 4 pages. The acknowledgments and references may start on an earlier page, if there is space.

% \ifinterspeechfinal
%      The Interspeech 2024 organisers
% \else
%      The authors
% \fi
% would like to thank ISCA and the organising committees of past Interspeech conferences for their help and for kindly providing the previous version of this template.

\bibliographystyle{IEEEtran}
\bibliography{mybib}

\end{document}